\documentclass{article}
\usepackage[a4paper]{geometry}
\usepackage{authblk}
\newgeometry {
    top = 2.5cm, 
    bottom = 2.5cm, 
    right = 2.5cm, 
    left = 2.5cm
}
\usepackage{authblk}
\usepackage{amsmath}
\usepackage{comment}
\usepackage{mathtools}
\usepackage{stmaryrd}
\usepackage{bm}
\usepackage{algorithm}
\usepackage{xcolor}
\usepackage{algpseudocode}
\usepackage{algorithmicx}
\usepackage{subcaption}
\usepackage[utf8]{inputenc} 
\usepackage[T1]{fontenc}    
\usepackage{hyperref}       
\usepackage{url}            
\usepackage{booktabs}       
\usepackage{amsfonts}       
\usepackage{nicefrac}       
\usepackage{microtype}      
\usepackage{lipsum}
\usepackage{fancyhdr}       
\usepackage{graphicx}       
\graphicspath{{media/}}     
\providecommand{\keywords}[1]
{
  \small	
  \textbf{\textit{Keywords---}} #1
}

\title{Accelerating high order discontinuous Galerkin solvers through a clustering-based viscous/turbulent-inviscid domain decomposition
}
\author[1]{Kheir-Eddine Otmani\thanks{Corresponding author: otmani.kheir-eddine@alumnos.upm.es}}
\author[1]{Andr\'es Mateo-Gab\'in}
\author[1,2]{Gonzalo Rubio}
\author[1,2]{Esteban Ferrer}

\affil[1]{ETSIAE-UPM-School of Aeronautics, Universidad Politécnica de Madrid, Plaza Cardenal Cisneros 3, E-28040 Madrid, Spain}
\affil[2]{Center for Computational Simulation, Universidad Politécnica de Madrid, Campus de Montegancedo, Boadilla del Monte, 28660 Madrid, Spain}
\begin{document}

\maketitle

\begin{abstract}
We explore the unsupervised clustering technique introduced in \cite{clustering_paper} to identify viscous/turbulent from inviscid regions in incompressible flows. The separation of regions allows solving the Navier-Stokes equations including Large Eddy Simulation closure models only in the viscous/turbulent ones, while solving the Euler equations in the remaining of the computational domain. By solving different sets of equations, the computational cost is significantly reduced.
This coupling strategy is implemented within a discontinuous Galerkin numerical framework, which allows discontinuous solutions (i.e., different set of equations) in neighbouring elements that interact through numerical fluxes.


The proposed strategy maintains the same accuracy at lower cost, when compared to solving the full Navier-Stokes equations throughout the entire domain. Validation of this approach is conducted across diverse flow regimes, spanning from unsteady laminar flows to  unsteady turbulent flows, including an airfoil section at Reynolds numbers $Re=10^3$ and $10^4$ and large angles of attack, 
and the flow past a wind turbine, modelled using actuator lines. 
The computational cost is reduced by $25\%$ and $29\%$ for the unsteady turbulent flow around an airfoil section and the flow past the wind turbine, 
respectively. 

In addition, to further accelerate the simulations, we combine the proposed decoupling with local $P$-adaptation, as proposed in \cite{adaptation_paper}. When doing so, we reduce the computational cost by $41\%$ and $45\%$ for the flow around the airfoil section and the flow past the wind turbine, 
respectively.  
\end{abstract}



\section{Introduction}

The Navier-Stokes equations describe the motion of fluids and include terms that account for convection and viscous diffusion. When supplemented with a turbulence subgrid model of Large Eddy Simulation (LES) type, the resulting equations are able to properly capture turbulence even in under-resolved, coarser meshes.
Large eddy simulations can be expensive and finding means to decrease the cost is important as this enables the study of more complex physics through larger and longer simulations.

One possibility to reduce the cost is to refine the mesh in the flow regions where viscous and turbulent effects are important. Indeed, viscous/turbulent regions require finer meshes to capture large gradients (e.g., boundary layers) and vortex dynamics (e.g., wakes). The increased resolution can be obtained by means of mesh or polynomial refinements when using high order methods \cite{KOMPENHANS2016216,KOMPENHANS201636,IMS2023112312}. 

A less explored alternative is to modify the equations such that these only include the terms that are relevant to the flow regions at hand. 
In regions where intense viscous effects are present (e.g., boundary layers, wakes), it is necessary to solve the full LES Navier-Stokes equations with high accuracy, while far from walls (and in the absence of free stream turbulence) it is possible to solve only the Euler equations (neglecting viscosity and turbulence). Domain decomposition ideas have been explored in the past. Zhang et al. \cite{IBLvis-inv2017} developed a non-parametric discontinuous Galerkin finite-element formulation for integral boundary layer equations, featuring strong viscous-inviscid coupling. In \cite{vis-invDG2022}, a high-order viscous-inviscid interaction solver for aerodynamic flows was introduced, employing a split formulation where viscous and inviscid effects are separately solved in overlapping domains near solid walls or the wake centerline. A coupled viscous-inviscid interaction scheme that integrates the continuity equation for potential flow with the three-dimensional integral boundary layer equations was presented in \cite{LOKATT201737}. To our knowledge, there has not been any work using clustering to separate viscous/turbulent from inviscid regions to accelerate high order solvers.
 
To apply any of the above mentioned techniques, it is  critical to be able to identify/separate the regions where viscous/turbulent effects are important from the rest. 
In this work, we explore the capability of the clustering methodology proposed in \cite{clustering_paper} to decompose the mesh and simplify the equations when possible (away from walls and wakes). The full Navier-Stokes equations are exclusively solved in regions where the viscous effects are prominent. Away from them (i.e. in inviscid regions) the viscous and turbulent terms are neglected. The challenge is to retain accuracy, as to when solving the full set of equations everywhere, while reducing the computational cost.



In recent years, clustering techniques based on machine learning have emerged as promising tools to address the challenge of flow region separation in CFD. Clustering algorithms group data points into distinct clusters based on their similarities, enabling the identification of  flow regions. These techniques leverage the power of data-driven approaches and can adapt to complex and dynamic flow patterns.
By applying clustering techniques to CFD simulations, we can automatically detect and separate flow regions with different characteristics. 
In the context of flow region detection, various researchers have used machine learning models and especially clustering techniques, see \cite{liUsingMachineLearning2020,saettaflowregionsml,mateogabin2023unsupervised,Callaham2021} and our previous work \cite{clustering_paper,adaptation_paper}. 

Here, we propose to use a clustering technique along with the feature space proposed in \cite{clustering_paper} to identify viscous-dominated rotational regions (including turbulent flow), using the principal invariants of strain and rotational rate tensors as inputs to the Gaussian Mixture Model (GMM) to detect/separate viscous/turbulent-dominated areas from the rest. Furthermore, in \cite{adaptation_paper}, we showed that the detected regions can be tracked during the simulation and can be used to dynamically adjust the local resolution used to approximate the solution ($P$-adaptation for high order methods) with the aim of reducing the computational cost while maintaining the level of accuracy. 

In this new work, the focus is to use the separated regions to simplify the equations to be solved, i.e., the full LES Navier-Stokes equations are to be solved only in the viscous/turbulent-dominated regions while the Euler equations are solved in the outer region. 
By doing so, we can concentrate the computational efforts in the regions where viscous/turbulent effects are important. The numerical method used to discretize the NS equations is the Discontinuous Galerkin Spectral Element Method (DGSEM) implemented in the open-source solver HORSES3D \cite{horses3d_paper}, since discontinuous Galerkin techniques allow to naturally handle discontinuous solutions. In the second part of the paper, we combine the proposed methodology (hybrid equations in the domain) with local $P$-adaptation to further accelerate the simulations. 

The work is organized as follows. We first introduce the methodology
in Section \ref{sec:method} including the viscous/turbulent region detection through a Gaussian mixture model, and details on the high order discontinuous Galerkin solver HORSES3D. Section \ref{hybrid_simulation} summarises the results obtained for the hybrid approach (solving LES-Navier-Stokes near walls and wakes and Euler in the rest) to show acceleration without loss of accuracy. In section \ref{hybrid_adapted_simulation}, we combine the new method with local $P$-adaptation to further accelerate the simulations. Finally, conclusions are summarised in section \ref{sec:conc}.  

\section{Methodology}\label{sec:method}
\subsection{Viscous/turbulent region detection through Gaussian mixture models} \label{sec:clustering_meth}
Distinct physics are present in different flow regions. For example, the flow near walls is characterized by important viscous effects, which lead to the development of boundary layers and within wakes, rotation, viscous effects and turbulence will be important. 
Far from the solid boundaries and the wake region, viscous effects, rotation/vorticity and turbulence in the flow are negligible.
To identify the viscous/turbulent flow regions, we proposed in \cite{clustering_paper} to use the principal invariants of the strain and rotational rate tensors as inputs to the Gaussian mixture unsupervised model to detect two different regions, a viscous, rotational/vortical region, on the one hand, and an inviscid outer region, on the other hand. The principal invariants of the strain and rotational rate tensor are defined as follows for incompressible flows:
\begin{align*} 
{Q_{S}}=\frac{1}{2}( tr((\boldsymbol { \mathbf S}))^2-tr(\boldsymbol { \mathbf S}^2))\;\; ; \;\;
 R_{S}=-\frac{1}{3}det(\boldsymbol { \mathbf S}),
\end{align*}
where $\boldsymbol {\mathbf S}$ is the strain rate tensor defined as $  \boldsymbol { \mathbf S}=\frac{1}{2}(\boldsymbol {\mathbf J}+\boldsymbol {\mathbf J}^T)$ and $ \boldsymbol{\mathbf J}=\nabla\boldsymbol{\mathbf U}$ is the gradient tensor of the velocity field $\boldsymbol{ \mathbf U}$. The rotational tensor $\boldsymbol { \mathbf \Omega}=\frac{1}{2}(\boldsymbol {\mathbf J}-\boldsymbol { \mathbf J}^T)$ has one invariant defined as: 
\begin{align*}
    Q_{\Omega}=-\frac{1}{2}tr( \boldsymbol {\mathbf \Omega}^2).
\end{align*}
The features $Q_S$,$R_S$ and $Q_\Omega$ will be used to train the Gaussian mixture model (GMM) to identify two distinct flow regions: a viscous/turbulent region and an inviscid outer region. The clustering will provide a node-wise partitioning of the computational domain, and made of Gauss-Legendre points that constitute the the Degrees of Freedom (DoF) in our high order numerical method. The GMM clustering provides each node with two probability memberships $p_{v}$ for the viscous/rotational region and $p_{i}$ for the inviscid outer region. These probabilities describe the model's estimation of the likelihood that a given node belongs to a particular region. To obtain an element-wise representation of the detected region, the mean of probability memberships belonging to each region will be computed inside each element as follows: 
\begin{equation}
  \bar{p}_v= \frac{1}{N}\sum_{j=1}^N {p_v}_j, \qquad {p_v}_j \in [0,1],
\end{equation}
\begin{equation}
  \bar{p}_i= \frac{1}{N}\sum_{j=1}^N {p_i}_j, \qquad {p_i}_j \in [0,1],
\end{equation}
where $N=(P+1)^{3}$ are the number of DoF of each element of the mesh, and $P$ denotes the polynomial order associated with the high-order discretization. An element will be assigned to the region with the highest mean of membership probabilities $\bar{p}=\max{(\bar{p}_i,\bar{p}_v)}$. The same process will be applied for all elements to supply each one with a region ID, viscous/turbulent or inviscid. 
\subsection{A high order discontinuous Galerkin solver}
All simulations have been performed using the high-order spectral element CFD solver, HORSES3D \cite{horses3d_paper}. Developed at ETSIAE–UPM (the School of Aeronautics of the Polytechnic University of Madrid), HORSES3D is a 3D parallel code designed for simulating fluid-flow phenomena. It employs the high-order discontinuous Galerkin spectral element method (DGSEM) and is implemented in modern Fortran 2003. The solver is adept at handling simulations governed by both compressible and incompressible Navier-Stokes equations and supports curvilinear hexahedral meshes of arbitrary order. The compressible Navier-Stokes equations together with details of the spatial DGSEM dicretisation are included in the appendix of this text.

HORSES3D stands out for its ability to handle anisotropic $p$-non-conforming elements, a key feature it offers. In our study, we exploit the adaptation capability of HORSES3D, adjusting the polynomial order uniformly within each element based on regions identified by the clustering technique. Most importantly, the solver incorporates a built-in Gaussian mixture model that is used to capture the viscous-dominated rotational regions during the run-time of the simulations, this built-in model allows the partitioning of the computational domain into two regions: viscous/turbulent and outer regions.

\subsection{Viscous/Turbulent-Inviscid  interactions} \label{clust_vis_inv}
The full LES Navier-Stokes equations incorporate convective, viscous diffusive and turbulent terms. Solving these equations in the entire computational domain can be expensive for flows that exhibit local features, such as in wakes of turbulent flows. In regions characterized by intense viscous effects, solving the full NS equations is warranted to accurately capture the intricate behavior. This ensures that viscous forces and rotational tendencies are properly accounted for a crucial understanding of boundary layer dynamics and near-wall interactions. Conversely, in regions where the flow is less influenced by viscous forces, an Euler equation, neglecting viscosity, can be solved. The Euler equation simplifies the computations by removing the need to account for viscous diffusion. 
The developed methodology exploits this behaviour and only computes viscous terms in the viscous domain, while in the inviscid domain, the viscous/turbulent terms will be neglected, which is equivalent to solving the Euler equation (inviscid flow). Let us note that the presented methodology affects only the viscous/turbulent terms in the governing equations, and therefore convective terms will be accounted for in the entire computational domain.  In HORSES3D, there are two types of element interfaces, interior faces where the face is shared between two neighbor elements, and boundary faces which are localized at the boundaries of the computational domain. 
For every element 
we can highlight three different situations that occur when partitioning the domain into viscous/turbulent-dominated and inviscid regions:
\begin{itemize}
\item Inviscid region interactions:
\\
Far from boundaries we find flow regions where viscous and turbulent effects are negligible. 
Here, we solve the Euler equations (neglecting the viscous and turbulent terms in the equation) for both elements and we only allow for the computation of the numerical convective flux to couple the elements. A Roe Riemann solver \cite{ROE1981357} is used as a numerical convective flux (but any other convective flux can be selected).  
\item Viscous and turbulent interactions:
\\
Near boundaries and within wakes, viscous and turbulent effects are important. 
In these elements, we solve the full LES Navier-Stokes equations for both elements and we couple the elements by computing both the numerical viscous and inviscid fluxes at the shared face. Similarly to the previous case, we use a Roe Riemann solver to compute the convective numerical flux and the BR1 viscous numerical flux \cite{BASSI1997267}. 
\item  Inviscid-Viscous/Turbulent interactions:
\\
At the edge of the clustering regions, we find elements that contain inside some degrees of freedom that are inviscid and some viscous/turbulent. In these cases, the mean probability memberships (described previously) is used to decide the final nature of the element. The connection between elements with different character is naturally handled by the discontinuous Galerkin method and the fluxes between elements. We remind the reader that the discontinuous Galerkin method allows for discontinuous solutions between neighbouring elements.

\end{itemize}
Two approaches to performing the clustering are studied: Static clustering, where the identification of the viscous/turbulent regions is performed only once in the simulation after the flow is fully developed, and dynamic clustering where the identification is performed every $n$ iterations, where $n$ should be selected a-priori. If $\Delta t$ is the time step used in the simulation then the clustering step is $\Delta c=\Delta t \times n$. 
First we use dynamic clustering in section \ref{hybrid_simulation} to develop the 
viscous/turbulent-inviscid coupling strategy. Then static clustering is used in section \ref{hybrid_adapted_simulation} where the viscous/turbulent-inviscid coupling strategy is complemented with the p-adaptation methodology. Both approaches yield similar levels of accuracy when detecting the regions, however dynamic clustering tends to be computationally more expensive than static clustering. In all the simulation shown in this work, time marching is conducted using a low-storage third order explicit Runge–Kutta scheme \cite{RK}. Finally, appendix \ref{staticVSdynamic} includes a comparison between these two approaches. 
\section{Results and discussion}
In this section, we first explore the advantages of hybrid simulations and then investigate the combination of hybrid simulations coupled with local polynomial adaptation.
\subsection{Hybrid simulations} \label{hybrid_simulation}
We apply the methodology to three distinct test cases, each increasing in complexity. Firstly, we demonstrate an unsteady laminar flow around a NACA0012 airfoil at $Re=10^3$, considering two different angles of attack: $\alpha=10^{\circ}$ and $\alpha=20^{\circ}$. Then, we analyze an unsteady turbulent flow around a NACA0012 airfoil at $Re=10^4$ with an angle of attack $\alpha=10^{\circ}$. Finally, we showcase the capabilities of the methodology with a flow past a wind turbine at $Re_c=103600$. The results achieved through the use of viscous/turbulent-inviscid coupling are denoted as "Hybrid HORSES3D", while the results obtained from solving the full Navier-Stokes equations are referred to as "Standard HORSES3D".

\subsubsection{Flow around a NACA0012 at $Re=10^3$}
\begin{figure}
\centering   		
\includegraphics[width=0.90\linewidth,height=5.00cm,keepaspectratio]{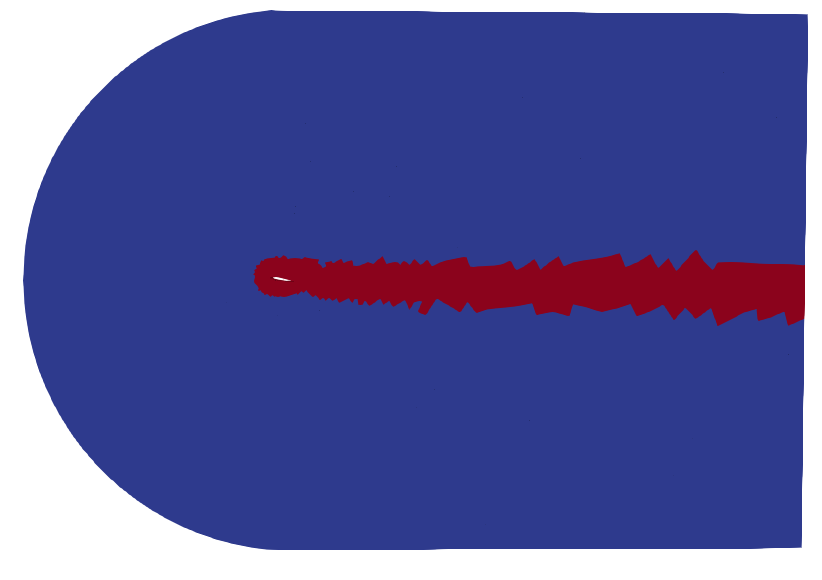}
\caption{Regions detected for the flow around a NACA0012 at $Re=10^3$, $\alpha=10^{\circ}$ at $t=42s$. \textcolor{red}{Red:} Viscous domain, \textcolor{blue}{Blue:} Inviscid domain.}
\label{regions_aoa10}
\end{figure}

\begin{figure}
\begin{subfigure}[b]{0.5\linewidth}
\centering   		\includegraphics[width=0.90\linewidth,height=5.00cm,keepaspectratio]{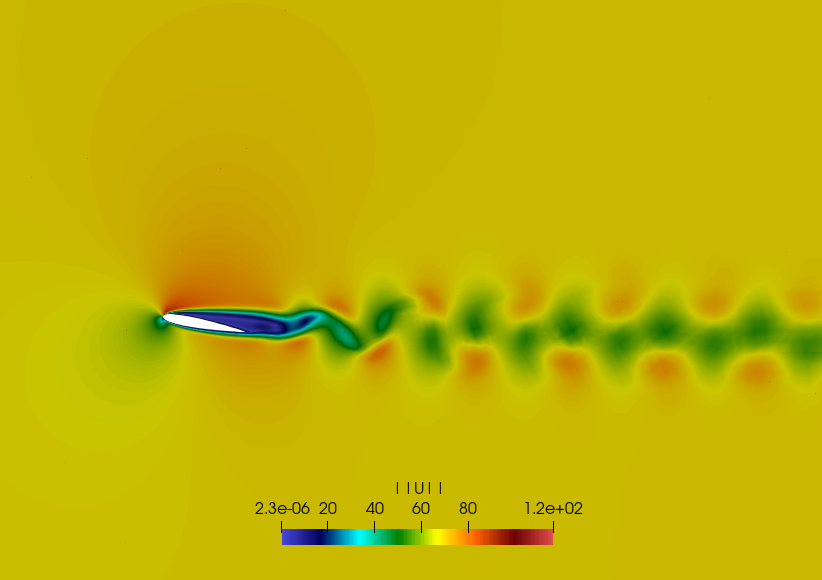}
\caption{Standard HORSES3D}
\label{naca_vel_standard}
     \end{subfigure}
\hfill
\begin{subfigure}[b]{0.5\linewidth}
\centering    \includegraphics[width=0.90\linewidth,height=5.00cm,keepaspectratio]{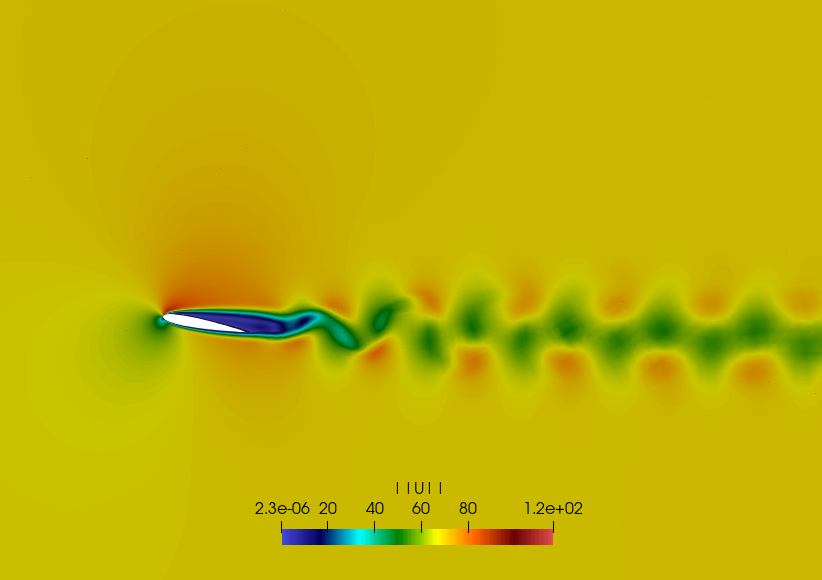}
    \caption{Hybrid HORSES3D}
\label{naca_vel_hybrid}    
\end{subfigure}
\caption{Velocity magnitude $||U||_2$ for the flow past an airfoil NACA0012 at $Re=10^3$, $\alpha=10^{\circ}$}
\label{naca_aoa10_vel}
\end{figure}

\begin{figure}
\begin{subfigure}[b]{0.5\linewidth}
\centering   		\includegraphics[width=0.90\linewidth,height=5.00cm,keepaspectratio]{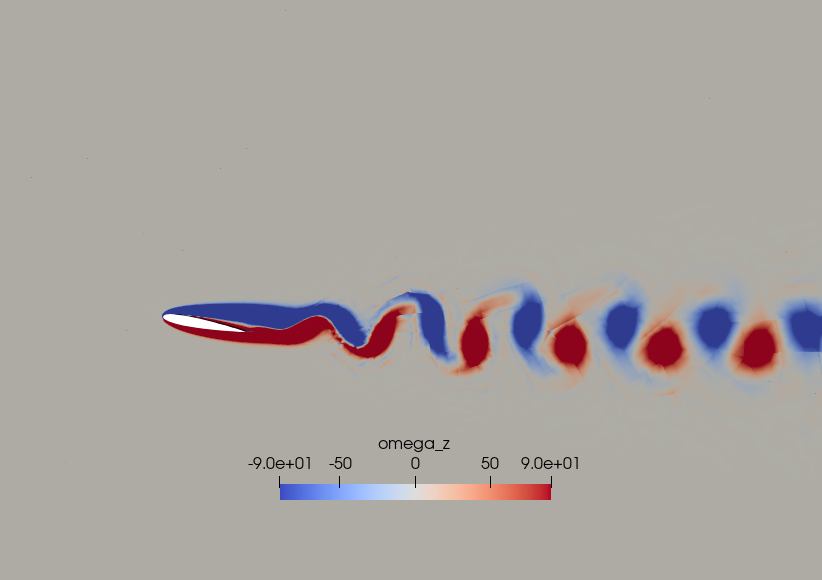}
\caption{Standard HORSES3D}
\label{naca_vort_standard}
     \end{subfigure}
\hfill
\begin{subfigure}[b]{0.5\linewidth}
\centering    \includegraphics[width=0.90\linewidth,height=5.00cm,keepaspectratio]{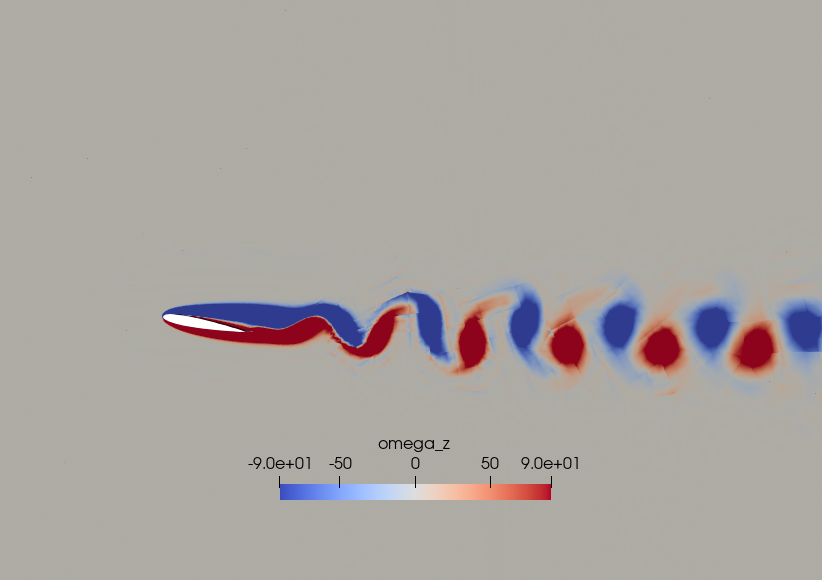}
    \caption{Hybrid HORSES3D}
\label{naca_vort_hybrid}    
\end{subfigure}
\caption{Spanwise vorticity $\omega_z$ for the flow past an airfoil NACA0012 at $Re=10^3$, $\alpha=10^{\circ}$}
\label{naca_aoa10_vort}
\end{figure}

\begin{figure}
\begin{subfigure}[b]{0.5\linewidth}
\centering   		\includegraphics[width=0.90\linewidth,height=5.00cm,keepaspectratio]{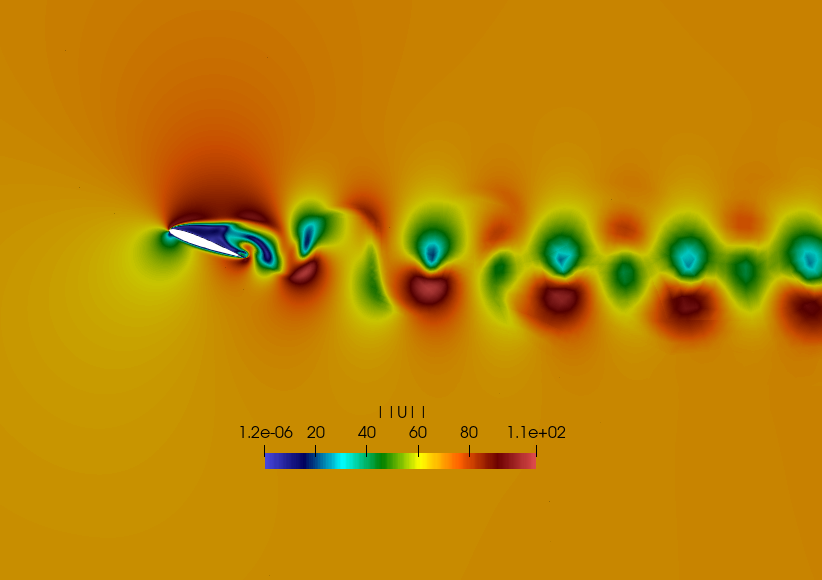}
\caption{Standard HORSES3D}
\label{naca20aoa_vel_standard}
     \end{subfigure}
\hfill
\begin{subfigure}[b]{0.5\linewidth}
\centering    \includegraphics[width=0.90\linewidth,height=5.00cm,keepaspectratio]{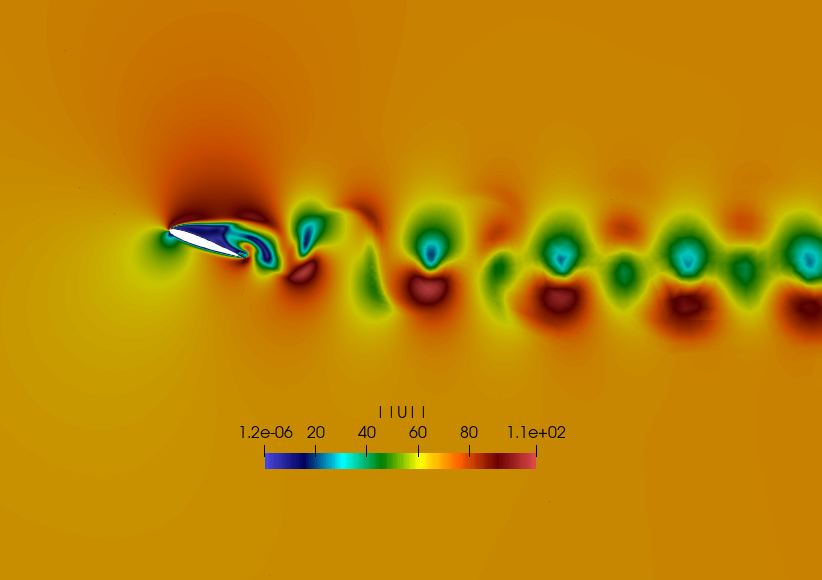}
    \caption{Hybrid HORSES3D}
\label{naca20aoa_vel_hybrid}    
\end{subfigure}
\caption{Velocity magnitude $||U||_2$ for the flow past an airfoil NACA0012 at $Re=10^3$, $\alpha=20^{\circ}$}
\label{naca_aoa20_vel}
\end{figure}

\begin{figure}
\begin{subfigure}[b]{0.5\linewidth}
\centering   		\includegraphics[width=0.90\linewidth,height=5.00cm,keepaspectratio]{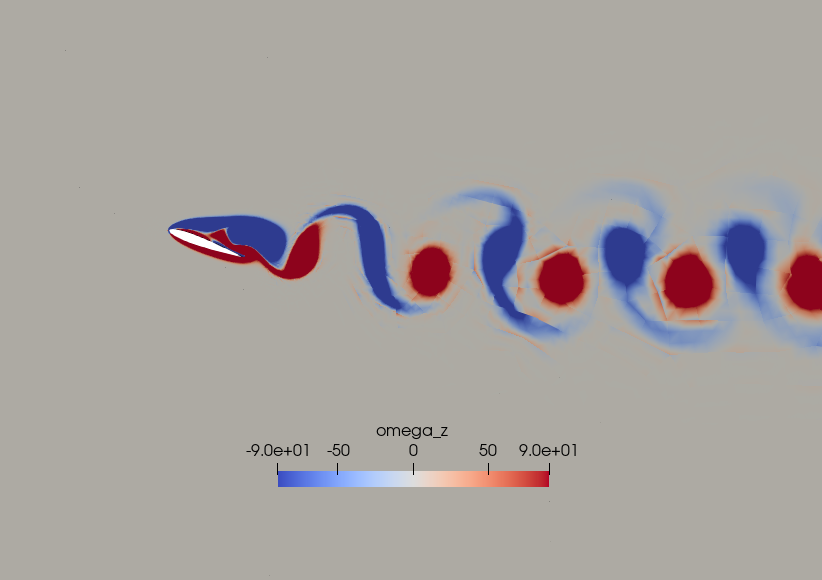}
\caption{Standard HORSES3D}
\label{naca20aoa_vort_standard}
     \end{subfigure}
\hfill
\begin{subfigure}[b]{0.5\linewidth}
\centering    \includegraphics[width=0.90\linewidth,height=5.00cm,keepaspectratio]{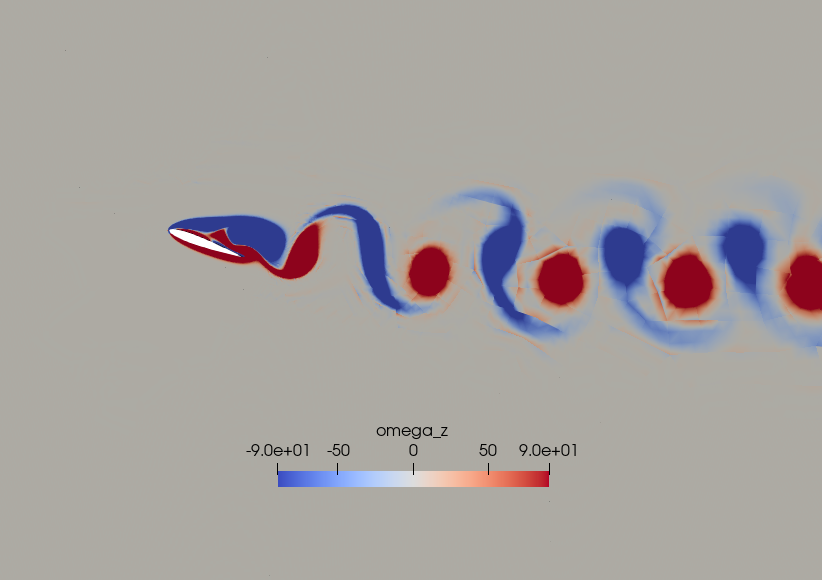}
    \caption{Hybrid HORSES3D}
\label{naca20aoa_vort_hybrid}    
\end{subfigure}
\caption{Spanwise vorticity $\omega_z$ for the flow past an airfoil NACA0012 at $Re=10^3$, $\alpha=20^{\circ}$}
\label{naca_aoa20_vort}
\end{figure}

The viscous/turbulent-inviscid coupling methodology is first validated with a two-dimensional flow around a NACA0012 airfoil at $Re=10^3$ at two different angles of attack  $\alpha= 10^{\circ}$ and $ \alpha = 20^{\circ}$. The mesh comprises $6039$ elements and a uniform polynomial order of $P=3$, resulting in a total of $193248$ degrees of freedom (when accounting for the Gauss-Legendre points in each element). At this Reynolds and angles of attack, an unsteady wake will develop.  We will challenge the capability of our methodology to track the regions of interest and solve the NS equations in the regions of high viscous effects as the angle of attack changes from $\alpha= 10^{\circ}$ to $\alpha= 20^{\circ}$. A non-dimensional time step of $\Delta t=6 \times 10^{-5}$ is used throughout all the simulations, and the clustering step for this case is $\Delta c=1.2 \times 10^{-3}$ resulting in performing the clustering every $20$ iterations during the simulation. The Mach number for this test case is set to $M=0.2$. Figure \ref{regions_aoa10} illustrates the viscous dominated regions detected at $t=42s$ for the flow around a NACA0012 at $Re=10^3$ with $\alpha=10^{\circ}$, showing that the clustering methodology, detailed in \ref{sec:method}, is able to detect the boundary layer and the wake of the airfoil where most of the viscous effects are concentrated. Further details on the clustering methodology, including other examples can be found in our previous works \cite{adaptation_paper,clustering_paper}.
To confirm the effectiveness of the viscous/turbulent-inviscid coupling methodology, the mean aerodynamic forces (mean drag $C_d$ and mean lift $C_l$ coefficients) and the Strouhal number St for $\alpha= 10^{\circ}$ and $\alpha= 20^{\circ}$  obtained with the "Hybrid HORSES3D" are compared in tables \ref{tab:naca_aoa10_re1000refvalues} and \ref{tab:naca_aoa20_re1000refvalues}, respectively, 
  with those obtained with the "Standard HORSES3D" as well as the results reported in previous studies of this test case.
  \begin{table}[h!]                           
	\centering
	\caption{Comparison of numerical results from the literature for the unsteady flow past an airfoil NACA0012 at $Re=10^3$ with $\alpha=10^{\circ}$.}
	\label{tab:naca_aoa10_re1000refvalues}
	\renewcommand{\arraystretch}{1.25}
	\begin{tabular}{cccc}
		\toprule
		 & $\mathrm{C}_{d}$ & $\mathrm{C}_{l}$ & St      \\
		\toprule
          Standard HORSES3D & 0.16744  &  0.41658 & 0.870 \\
          Hybrid HORSES3D & 0.16737 &  0.41631 & 0.869 \\
		  Kouser et al. \cite{krouserNACA}& 0.16608 & 0.41836  & 0.876 \\
		Di Ilio et al. \cite{DIILIO2018200} & 0.15652 & 0.41470 & - \\
		 Kurtulus \cite{kurtulusnaca} & 0.16304 & 0.42058 & - \\
		\bottomrule
	\end{tabular}
\end{table} 
The average mean drag and lift coefficients are computed within an interval of time $T=24U_{\infty}/c$, where $U_{\infty}$ is the free-stream velocity and $c$ is the airfoil chord. The mean drag and lift coefficients along with the Strouhal number predicted by the "Hybrid HORSES3D" for this test case exhibit close agreement with those obtained using the "Standard HORSES3D" and are consistent with findings from previous studies. 
\begin{table}[h!]                           
	\centering
	\caption{Comparison of numerical results from the literature for the unsteady flow past an airfoil NACA0012 at $Re=10^3$ with $\alpha=20^{\circ}$.}
	\label{tab:naca_aoa20_re1000refvalues}
	\renewcommand{\arraystretch}{1.25}
	\begin{tabular}{cccc}
		\toprule
		 & $\mathrm{C}_{d}$ & $\mathrm{C}_{l}$ & St      \\
		\toprule
          Standard HORSES3D & 0.45200  &  0.91200  & 0.539 \\
          Hybrid HORSES3D & 0.45196 &  0.91182  & 0.539 \\
		  Kouser et al. \cite{krouserNACA}& 0.44595 & 0.92811  & 0.531 \\
		Di Ilio et al. \cite{DIILIO2018200} & 0.44705 & 0.90666 & - \\
		 Kurtulus \cite{kurtulusnaca} & 0.44117 & 0.89066 & - \\
		\bottomrule
	\end{tabular}
\end{table}
In figures \ref{naca_aoa10_vel}, \ref{naca_aoa10_vort}, \ref{naca_aoa20_vel} and \ref{naca_aoa20_vort}, we compare the magnitude velocity $||U||_2$ and the spanwise vorticity $\omega_z$ for "Standard HORSES3D" and "Hybrid HORSES3D" for $\alpha=\{10^{\circ},20^{\circ}\}$.
There are no observable differences in the velocity and vorticity profiles between the "Hybrid HORSES3D" and "Standard HORSES3D" simulations, implying that the employment of the viscous/turbulent-inviscid coupling methodology could potentially yield comparable accuracy to the "Standard HORSES3D" simulation, which solves the Navier-Stokes equations across the entire computational domain. The "Hybrid HORSES3D" solution was obtained in $14 \%$ and $16 \%$ less computational time than the "Standard HORSES3D" solution, for the angles of attack $\alpha=10^{\circ}$ and $\alpha=20^{\circ}$, respectively as shown in table \ref{comp_time_naca0012_1000} .
\begin{table}[h!]                           
	\centering
	\caption{Comparison of computational time of the Hybrid and Standard HORSES3D simulations for the flow around a NACA0012 at $Re=10^3$ with a uniform polynomial order $P=3$.} 
        \label{comp_time_naca0012_1000}
        \begin{tabular}
        {p{28mm}|p{25mm} p{25mm}p{25mm}}
        \hline
Test case&Simulation type& Compt.  time (s)& Reduction of Compt.time \\        
\hline
NACA0012 $Re=10^{3},\ \  \alpha=10^{\circ}$&Standard HORSES3D&$1.503 \times 10^5$&-\\
\cline{2-4}
&Hybrid HORSES3D&$1.298 \times 10^5$& $14 \%$\\
\hline
NACA0012 $Re=10^{3} ,\ \ \alpha=20^{\circ}$&Standard HORSES3D&$1.554 \times 10^5$&-\\
\cline{2-4}
&Hybrid HORSES3D&$1.319 \times 10^5$&$16\%$\\
\hline
\end{tabular}
	
\end{table}

 \subsubsection{Flow around a NACA0012 at $Re=10^4$} \label{NACA0012_10000_hyb}
We now challenge the methodology with the unsteady turbulent three-dimensional flow around a NACA0012 at $Re=10^4$ and an angle of attack $\alpha=10^{\circ}$. This test case has been studied by various researchers \cite{ferrer2017interior,ZHOU2011329NACA10000}. Now, we consider a mesh comprised of $81420$ elements and a uniform polynomial order $P=3$ resulting in a total of degrees of freedom $DoF=5.1\times10^6$. A total of 10 elements have been used to extrude the mesh in the spanwise direction with a length $L_{z}/c=1$, where $c$ is the chord of the airfoil. A large eddy simulation with a Vreman sub-grid closure turbulence model (see Appendix of this text for details) and a Mach number $M=0.2$ has been conducted to show the effectiveness of the viscous/turbulent-inviscid coupling methodology. 

\begin{figure}
\begin{subfigure}[b]{0.5\linewidth}
\centering   		\includegraphics[width=0.9\linewidth,height=5.00cm,keepaspectratio]{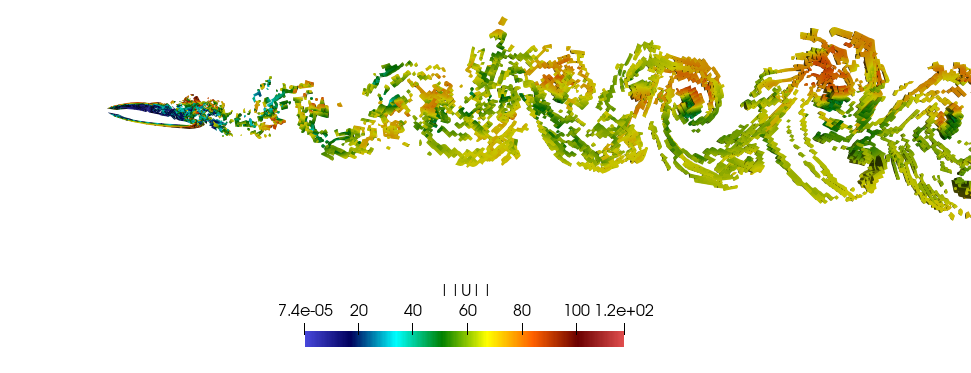}
\caption{Standard HORSES3D}
\label{naca10000_Qcrit_standard}
     \end{subfigure}
\hfill
\begin{subfigure}[b]{0.5\linewidth}
\centering    \includegraphics[width=0.90\linewidth,height=5.00cm,keepaspectratio]{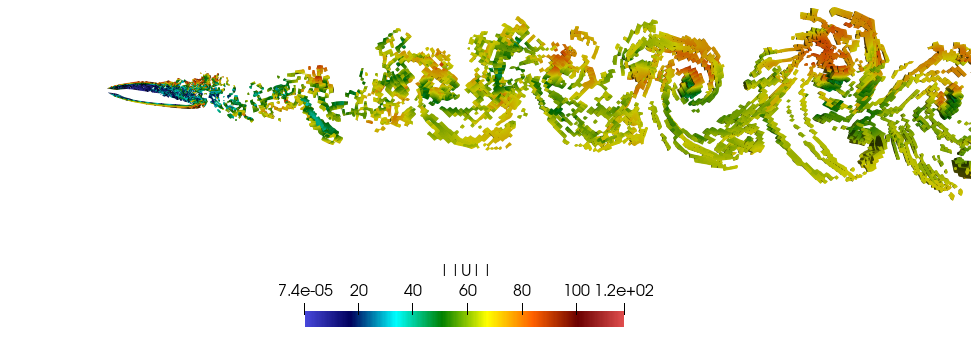}
    \caption{Hybrid HORSES3D}
\label{naca10000_Qcrit_hybrid}    
\end{subfigure}
\caption{$Q$-criterion contours, colored with L2 norm of velocity field $||U||_2$ for the flow around an airfoil NACA0012 at $Re=10^4$, $\alpha=10^{\circ}$}
\label{naca10000_Qcrit}
\end{figure}

Both simulations "Standard HORSES3D" and "Hybrid HORSES3D" were conducted with a non-dimensional time step $\Delta t=5\times10^{-5}$. The clustering for this test case is made every 20 iterations which is equivalent to a clustering step $\Delta c=10^{-3}$ as explained in section \ref{sec:method}. The drag and lift coefficients are averaged within an interval of time $T=15U_\infty/c$. The obtained results of "Standard HORSES3D" and "Hybrid HORSES3D" simulations are presented in table \ref{tab:naca_aoa10_re10000refvalues}, and compared against 
experimental data \cite{ZHOU2011329NACA10000,sunada1997,wang2014}. The drag coefficient obtained from "Hybrid HORSES3D" closely matches that of "Standard HORSES3D" with an error of $7 \times 10^{-4}$, and falls well in line with the results reported in the literature. The lift coefficient calculated using "Hybrid HORSES3D" aligns closely with the one predicted by "Standard HORSES3D". As pointed out by Ferrer \cite{ferrer2017interior}, the mean lift predictions for this test case at this angle of attack are highly sensitive to the testing conditions (e.g., Reynolds number, turbulence intensity) which explains the scattering of results for experiments at comparable Reynolds numbers.
In figure \ref{naca10000_Qcrit}, the $Q$-criterion contours colored with velocity field magnitude are shown for both "Hybrid" and "Standard" HORSES3D. Comparing figures \ref{naca10000_Qcrit_hybrid} and \ref{naca10000_Qcrit_standard} reveals that the viscous/turbulent-inviscid coupling yields a solution with the same level of accuracy as the "Standard HORSES3D" solution. When applying the proposed methodology, we could reduce the computational cost by $25 \%$, as summarised in table \ref{comp_time_naca0012_10000}.
\begin{table}[h!]                           
	\centering
	\caption{Comparison of numerical and experimental results from the literature for the unsteady turbulent flow past an airfoil NACA0012 at $Re=10^4$ with $\alpha=10^{\circ}$.}
	\label{tab:naca_aoa10_re10000refvalues}
	\renewcommand{\arraystretch}{1.25}
	\begin{tabular}{cccc}
		\toprule
		 & $Re$ & $\mathrm{C}_{d}$  & $\mathrm{C}_{l}$     \\
		\toprule
          Standard HORSES3D & $10^4$  &  0.14586  & 0.55929 \\
          Hybrid HORSES3D & $10^4$ & 0.14513  & 0.56466 \\
		 Exp: Zhou et al. \cite{ZHOU2011329NACA10000}& $1.05 \times 10^4$ & 0.13608  & 0.61043 \\
          Exp: Sunada et al. \cite{sunada1997} & $4\times10^3$ &  0.14262 & 0.36395 \\
          Exp: Wang et al. \cite{wang2014} & $5.30 \times 10^3$ & 0.14836 & 0.48233 \\
		\bottomrule
	\end{tabular}
\end{table}
\begin{table}[h!]                           
	\centering
	\caption{Comparison of computational time of the Hybrid and Standard HORSES3D simulations for the flow around a NACA0012 at $Re=10^4$ with a uniform polynomial order $P=3$.}
        \label{comp_time_naca0012_10000}
        \begin{tabular}{p{25mm}|p{25mm}p{25mm}p{25mm}}
        \hline
Test case&Simulation type&Compt.time (s)& Reduction of Compt. time \\        
\hline
NACA0012 $Re=10^{4},\ \  \alpha=10^{\circ}$&Standard HORSES3D&$3.23 \times 10^6$&-\\
\cline{2-4}
&Hybrid HORSES3D&$2.422 \times 10^6$& $25 \%$\\
\hline
\end{tabular}
	
\end{table}
\subsubsection{Flow past a wind turbine
}\label{wind_turbine}
\begin{figure}[h!]
  \centering
  \includegraphics[width=\textwidth]{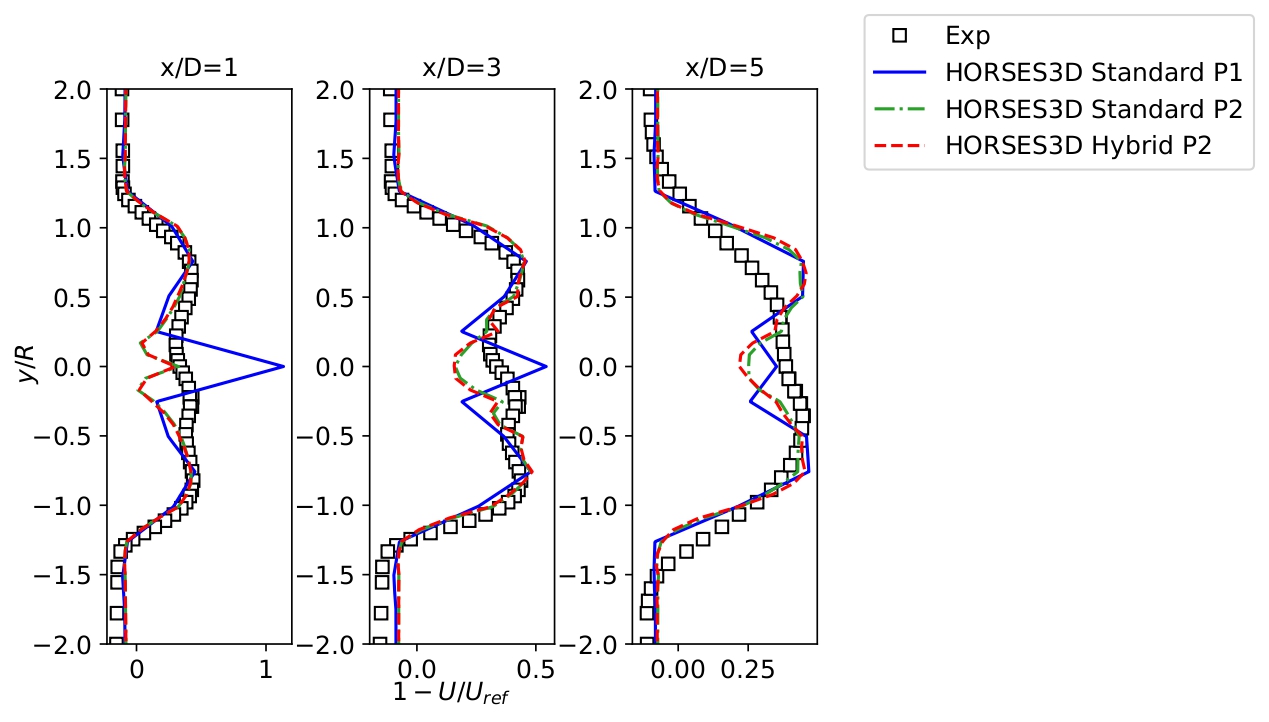} 
  \caption{Horizontal profiles of mean streamwise velocity deficit at three downstream positions $x/D$ for "Standard HORSES3D" $P=1$ and $P=2$ and "Hybrid HORSES3D" $P=2$ 
  compared against 
  experimental data (Exp) \cite{krogstad2013blind}.}
  \label{fig:turbine_comparison_exp_hybrid} 
\end{figure}

\begin{figure*}[!h]
    	\begin{subfigure}[h!]{0.5\linewidth}
    		\centering
			\includegraphics[width=1.0\linewidth]{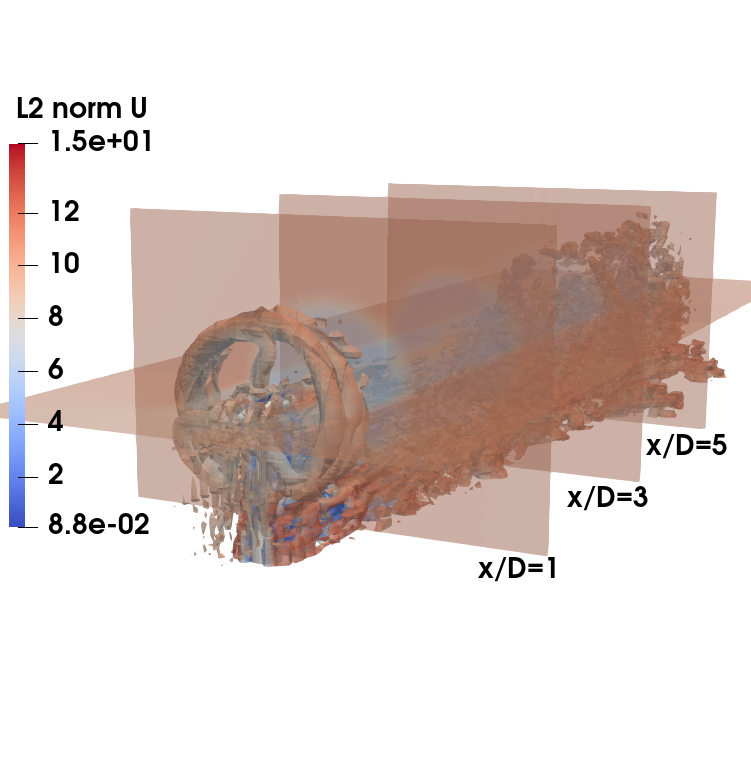}
    \caption{Standard HORSES3D}
    		\label{fig:turbine_contour_standard}
    	\end{subfigure}%
    	     \hfill
    	\begin{subfigure}[h!]{0.5\linewidth}
			\includegraphics[width=1.0\linewidth]{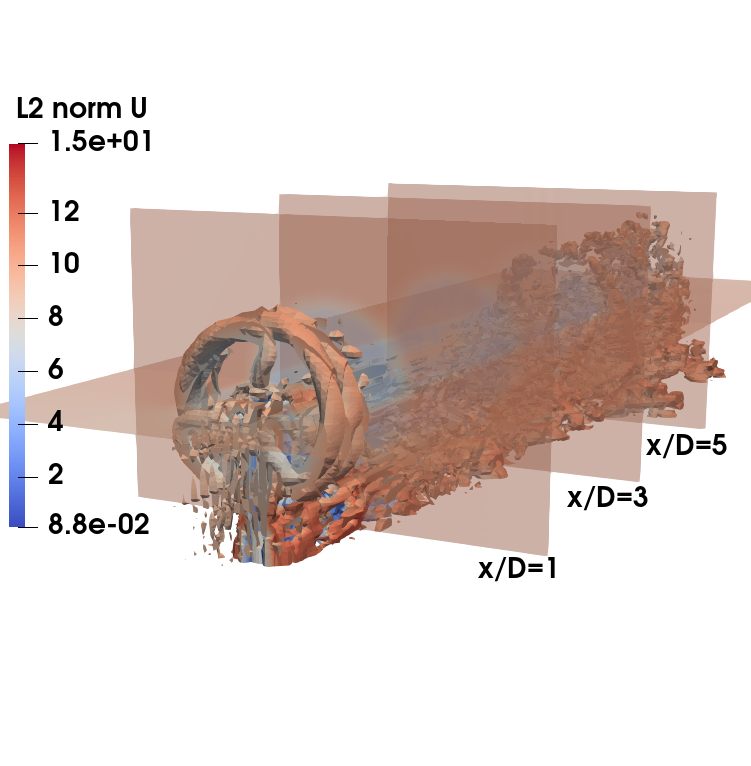}
    \caption{Hybrid HORSES3D}
    		\label{fig:turbine_contour_hybrid}
    	\end{subfigure}
    	\caption{ Contours of the Q-criterion, with colors representing the L2 norm of the velocity, for the flow past a wind turbine.}
    	\label{fig:turbine_Qcrit_vis_inv}
    \end{figure*}

The viscous/turbulent-inviscid coupling strategy is now applied to the three-dimensional turbulent flow past a wind turbine at $Re_c=103600$ (based on the blade chord). This flow has been tested experimentally at the Norwegian University of Science and Technology, the wind turbine has a diameter $D=0.894$m, the blades are made up of NREL S826 airfoils \cite{krogstad2013blind}. The blind test used a wind tunnel of dimensions $[L \times W \times H] = [11.15 \times 2.71 \times 1.8]$ m, a low turbulent intensity of $0.3 \%$, and a uniform inflow velocity. Various tip speed ratios ($\delta$) were used in the blind test. In this study, we use the optimal tip speed ratio $\delta=\gamma D/2U_{\infty} = 6$, with $U_{\infty}=10$~m/s and $\gamma=134.228$~rad/s. The blade tip Reynolds number for this case is $Re_c = \delta U_{\infty} c_{tip}/\nu= 103600$, where $c_{tip}=0.025926$ m is the tip chord length and $\nu$ is the kinematic viscosity of air. An immersed boundary method \cite{KOU2022110721,horses3d_paper} has been used to model the tower and the nacelle, the rotating blades were modeled using an actuator line method \cite{sorensen2002numerical}. Two large eddy simulations with a Vreman sub-grid closure model and polynomial order $P\in\{1,2\}$ have been conducted using the HORSES3D numerical framework \cite{horses3d_paper}. A Cartesian mesh is generated with the same size as the wind tunnel, consisting of $[128 \times 24 \times 24]$ elements. This results in a $D/\Delta x$ ratio of approximately 10, being $D$ the diameter of the turbine. It is important to note that, since we use a high-order method, the spatial resolution is also increased by raising the polynomial order. The Mach number is set to $M \approx 0.03$, a free slip boundary condition is used in the wind tunnel walls and no inlet turbulence is used. The total number of degrees of freedom when considering $P=2$, is $DoF=1.99 \times 10^6$. We run this test case using a polynomial order $P=1$ and $P=2$ for the "Standard HORSES3D" simulations, while we only use $P=2$ for the "Hybrid HORSES3D" simulation. A non-dimensional time step of $\Delta t = 1.6 \times 10^{-4}$ has been used for all the simulations. For the "Hybrid HORSES3D" simulation, the clustering step is $\Delta c \approx 3.34 \times 10^{-3}$. The simulations are conducted for $T=2\,s$. After that, the statistics are then gathered within an interval of length of $0.5\,s$. Figure \ref{fig:turbine_comparison_exp_hybrid} depicts the horizontal profiles of time-averaged streamwise velocity deficit at different downstream positions for the "Standard HORSES3D" and "Hybrid HORSES3D". These findings are compared with experimental data (Exp) \cite{krogstad2013blind}. 
We can observe a close agreement between the "Hybrid HORSES3D P2"  results and the "Standard HORSESD3D P2" simulation suggesting that the viscous/turbulent-inviscid coupling yields a very similar level of accuracy, when compared to "Standard HORSES3D". The differences observed between the numerical results and the experimental data can be attributed, in part, to the requirement for increased resolutions ($P>2$) and the utilization of more sophisticated blade modeling techniques (such as the incorporation of sliding meshes). The contours of the $Q$-criterion for "Hybrid HORSES3D P2" and "Standard HORSESD3D P2" are shown in figure \ref{fig:turbine_Qcrit_vis_inv}.
We can see that the "Hybrid HORSES3D" simulation is able to capture the vortical structures in the the wake of the turbine with a similar level of accuracy compared to the "Standard HORSES3D P2", proving that the viscous/turbulent-inviscid strategy was able to recreate a solution as accurate as when solving the full NS+LES equations in the entire domain while reducing the computational cost by $29 \%$ as shown in table \ref{comp_time_turbinehyb}. Finally, the table also includes the error in the outer inviscid region (relative to a P3 solution) for the Standard and Hybrid approaches. It can be seen that the errors are very similar, showing that no additional errors are included when hybridizing the simulations.
\begin{table}[h!]                           
	\centering
	\caption{Comparison of computational time of the Hybrid and Standard HORSES3D simulations for the flow past a wind turbine.}
        \label{comp_time_turbinehyb}
        \begin{tabular}{p{20mm}|p{25mm}p{15mm} p{20mm}p{20mm}p{20mm}}
        \hline
Test case&Simulation type&Polynomial order& Compt. time (s)& Reduction of Compt.time& $|| U_{P_3}- U_{P_x}||_{2}$ \\        
\hline
Wind turbine at $Re_c=103600$&Standard HORSES3D&$2$&$6.25 \times 10^5$&-&$0.10$\\
\cline{2-6}
&Standard HORSES3D&$1$&$3.508 \times 10^5$& $54 \%$&$1.30$\\
\cline{2-6}
&Hybrid HORSES3D&$2$&$4.44 \times 10^5$& $29 \%$&$0.11$\\
\hline
\end{tabular}
\end{table}

\subsection{Hybrid and $P$-adapted simulations} \label{hybrid_adapted_simulation}
The proposed methodology has been applied successfully to different flow regimes showing accelerations in all cases. 
In this section, we will combine the hybrid approach with $P$-adaptation. In the inviscid region, we will reduce the polynomial order while only solving the Euler equations, with the aim of further accelerating the simulations while preserving the accuracy.

Dynamic clustering has been employed to identify the viscous/turbulent regions during run-time simulations. This approach is effective when the targeted regions for detection change significantly with time. When the flow exhibits time-periodic behavior, such as in flows characterized by finite number of vortex shedding frequencies, static clustering can be used. In static clustering, the flow regions detection is performed only once in the simulation, and a representative clustering of the regions can be obtained within a vortex shedding cycle, static clustering is often preferred since it reduces the computational overhead that may result from performing the clustering dynamically during run-time simulations. A comparison between both approaches is provided in Appendix \ref{staticVSdynamic}. There, we include a comparison between dynamic and static clustering and report computational times for both, highlighting the advantage of static clustering.
In our previous work \cite{adaptation_paper}, we used static clustering for $P$-adaptation and it has been shown that once the flow is fully developed, a single snapshot can be used to cluster the flow regions. 
In this section and with the aim of obtaining optimal performance, we select static clustering to detect the viscous/turbulent regions.
We again consider 
the flow around a NACA0012 at $Re=10^4$ with angle of attack $\alpha=10^{\circ}$ and the flow past a turbine. In the text, we refer to the combination of different sets of equations together with adaptation as: "Hybrid adapted HORSES3D" while the results obtained using uniform polynomial order as well as solving the full NS equations in the entire domain are denoted as: "Standard HORSES3D".
\subsubsection{Flow around a NACA0012 at $Re=10^4$}
\begin{figure}
    \centering
  \includegraphics[width=0.9\linewidth,height=5.00cm,keepaspectratio]{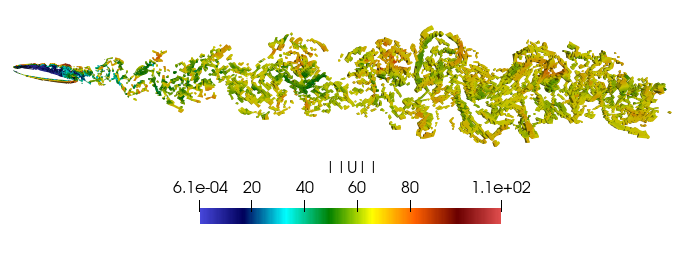}
    \caption{$Q$-criterion contours, colored with L2 norm of velocity field $||U||_2$, for “Hybrid adapted HORSES3D” simulation of flow around NACA0012 airfoil at $Re=10^4$, $\alpha=10^\circ$.}
    \label{naca1000_Qcrit_hyb_adapt_static}
\end{figure}

Following the same setup described in \ref{NACA0012_10000_hyb}, we conduct the "Standard HORSES3D" simulation with a uniform polynomial order $P=3$. For the "Hybrid adapted HORSES3D" simulation, we run the simulation until the flow is fully developed using $P_{init}=3$, we perform the clustering 
to detect the viscous/turbulent regions in the flow field.
For this simulation, we set $P_{cluster}=3$ and $P_{inviscid}=1$ and we restart the simulation until reaching a statistical convergence state. The diffusive and turbulent terms are accounted for only in the elements assigned to the viscous domain using the viscous/turbulent-inviscid coupling strategy presented in \ref{clust_vis_inv}. In table \ref{tab:naca_aoa10_re10000refvalues_hyb_adapted}, we present the mean drag and lift coefficients predicted by the "Standard HORSES3D" and "Hybrid adapted HORSES3D", these results are compared against previous experimental and numerical studies. The mean drag predictions of "Hybrid adapted HORSES3D" match closely the ones predicted by "Standard HORSES3D".
\begin{table}[h!]                           
	\centering
	\caption{Comparison of numerical and experimental results from the literature for the unsteady turbulent flow past an airfoil NACA0012 at $Re=10^4$ with $\alpha=10^{\circ}$.}
	\label{tab:naca_aoa10_re10000refvalues_hyb_adapted}
	\renewcommand{\arraystretch}{1.25}
	\begin{tabular}{cccc}
		\toprule
		 & $Re$ & $\mathrm{C}_{d}$  & $\mathrm{C}_{l}$     \\
		\toprule
          Standard HORSES3D & $10^4$  &  0.14586  & 0.55929 \\
          Hybrid Adapted HORSES3D & $10^4$ & 0.14206  & 0.54822 \\
		 Exp: Zhou et al. \cite{ZHOU2011329NACA10000}& $1.05 \times 10^4$ & 0.13608  & 0.61043 \\
          Exp: Sunada et al. \cite{sunada1997} & $4\times10^3$ &  0.14262 & 0.36395 \\
          Exp: Wang et al. \cite{wang2014} & $5.30 \times 10^3$ & 0.14836 & 0.48233 \\
		\bottomrule
	\end{tabular}
\end{table}
In section \ref{NACA0012_10000_hyb}, we discussed how the mean lift predictions for this specific test case can be highly affected by the testing conditions. This is the reason why the experimental data shows a considerable amount of variability, which can explain the variations observed in the predictions when the Reynolds number is changed. A visualization of the $Q$-criterion colored with velocity magnitude for the "Hybrid adapted HORSES3D" is provided in figure \ref{naca1000_Qcrit_hyb_adapt_static}.
In table \ref{comp_time_NACA0012_LES_static}, we present the computational time, the reduction in the number of Degrees of Freedom (DoF), and the reduction in computational cost achieved when using the “Hybrid adapted HORSES3D” strategy.
\begin{table}[h!]                           
	\centering
	\caption{Comparison of computational time of the Hybrid Adapted and Standard HORSES3D simulations for the flow past a NACA0012 at $Re_c=10^4$}
        \label{comp_time_NACA0012_LES_static}
        \begin{tabular}{p{20mm}|p{20mm}p{13mm} p{13mm}p{20mm}p{20mm}p{20mm}}
        \hline
Test case&Simulation type&$P_{cluster}$&$P_{inviscid}$& Compt.time (s)& Reduction of DoF &  Reduction of Compt.time \\        
\hline
NACA0012 at $Re_c=10^4$&Standard HORSES3D&$3$&$3$&$3.23 \times 10^6$&-&-\\
\cline{2-7}
&Hybrid Adapted HORSES3D&$3$&$1$&$1.923 \times 10^6$& $51 \%$&$41 \%$\\
\hline
\end{tabular}
\end{table}
When using the "Hybrid adapted HORSES3D" methodology, the number of DoF is reduced by $51\%$ resulting in a mean polynomial order $P=1.88$, and a reduction in computational cost of $41\%$, when compared to "Standard HORSES3D" simulation. 
\subsubsection{Flow past a wind turbine} 
\begin{figure}[h!]
  \centering
  \includegraphics[width=1\textwidth]{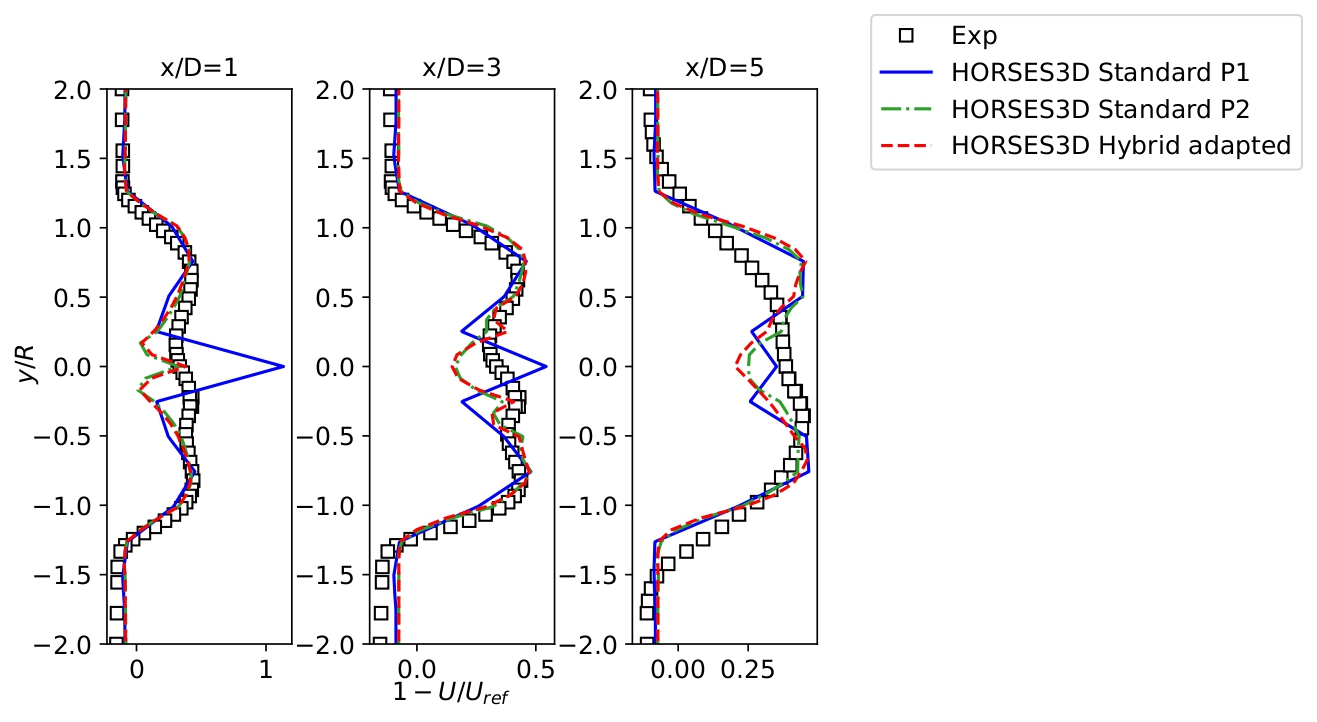} 
  \caption{Horizontal profiles of mean streamwise velocity deficit at three downstream positions $x/D$ for "Standard HORSES3D" $P=1$ and $P=2$ and "Hybrid Adapted HORSES3D" $P_{cluster}=2, P_{inviscid}=1$ compared against experimental data (Exp) \cite{krogstad2013blind}.}
  \label{fig:turbine_comparison_exp_hybrid_adapted} 
\end{figure}

\begin{figure}[h!]
  \centering
  \includegraphics[width=0.7\textwidth]{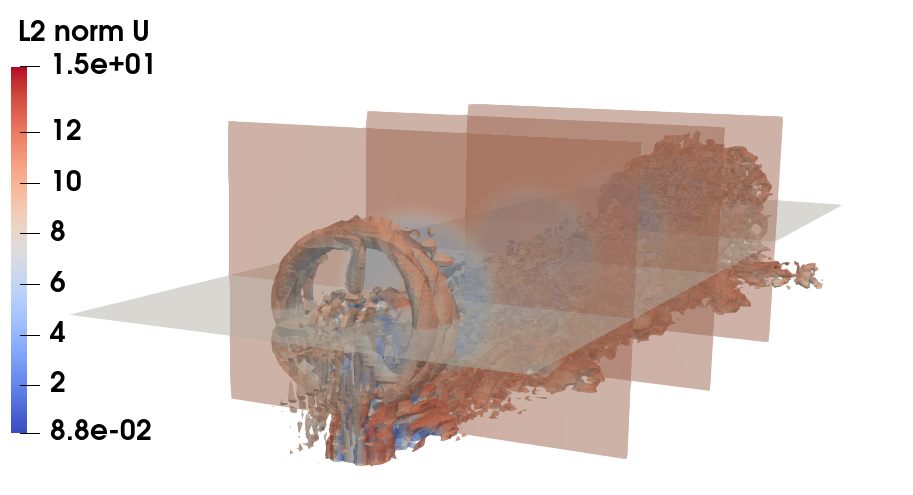} 
  \caption{Contours of the Q-criterion, with colors representing the L2 norm of the velocity field, for the "Hybrid adapted HORSES3D" flow past a wind turbine.}
  \label{fig:turbine_Qcrit_hybrid_adapted} 
\end{figure}

To simulate this case, we utilized the setup described in \ref{wind_turbine}, and run the simulations for 2 seconds with $P_{init} \in \{1,2\}$. For the hybrid adapted simulation, we set $P=2$ until the wake is fully developed, the clustering is then performed to obtain a partitioning of the domain. 
The simulation is restarted using the polynomial order distribution $P$ with $P_{cluster}=2$ and $P_{inviscid}=1$, we continue the simulation for an additional 1 second, but only updating the viscous/turbulent terms in the elements belonging to the viscous region, following the same approach proposed in \cite{adaptation_paper}. Finally, the statistics are collected within an interval of $0.5$ seconds. As in \ref{wind_turbine}, the horizontal profiles of mean streamwise velocity at the positions $x/D=1,3,5$ downstream the turbine are illustrated in figure \ref{fig:turbine_comparison_exp_hybrid_adapted}.
The "Hybrid adapted HORSES3D" showed close agreement with those of the "Standard HORSES3D P2", proving that the combination of the hybrid approach with local $P$-adaptation does not damage the accuracy of the solver. 
The new approach was able to reduce the number of DoF by $46 \%$ and the computational cost by $45 \%$.
The $Q$-criterion contours colored with the velocity magnitude for "Hybrid adapted HORSES3D" are depicted in figure \ref{fig:turbine_Qcrit_hybrid_adapted}.

A computational time reduction of $16\%$ has been achieved in obtaining the numerical solution for "Hybrid Adapted HORSES3D", when compared to the computational time for "Hybrid HORSES3D P2" reported in table~\ref{comp_time_turbinehyb}.    
\begin{table}[h!]                           
	\centering
	\caption{Comparison of computational time of the Hybrid Adapted and Standard HORSES3D simulations for the flow past a wind turbine}
        \label{comp_time_turbine_static}
        \begin{tabular}{p{15mm}|p{20mm}p{15mm} p{15mm}p{20mm}p{20mm}p{20mm}}
        \hline
Test case&Simulation type&$P_{cluster}$&$P_{inviscid}$& Compt.time (s)& Reduction of DoF &  Reduction of Compt. time \\        
\hline
Wind turbine at $Re_c=103600$&Standard HORSES3D&$2$&$2$&$6.497 \times 10^5$&-&-\\
\cline{2-7}
&Standard HORSES3D&$1$&$1$&$3.508 \times 10^5$& $70 \%$&$54 \%$\\
\cline{2-7}
&Hybrid Adapted HORSES3D&$2$&$1$&$3.58 \times 10^5$& $46 \%$&$45 \%$\\
\hline
\end{tabular}
\end{table}
\section{Conclusion}\label{sec:conc}
We have proposed a novel clustering-based viscous/turbulent-inviscid coupling strategy to accelerate high order discontinuous Galerkin solvers. This methodology can achieve a similar level of numerical accuracy as solving the full LES NS in the entire computational domain. The element-wise treatment of the DG numerical framework facilitates the implementation of such a strategy. We have shown that for solving different sets of equations (NS and Euler equations), one has to care only about the viscous numerical flux used to couple the elements at the interface between both regions. Keeping this numerical viscous flux single-valued ensures consistency and conservative properties of the scheme. By applying the viscous/turbulent-inviscid strategy, we could reduce the complexity of the governing equations in the inviscid-outer region which results in reducing the computational cost of the considered simulations, remarkably, the computational times of the NACA0012 at $Re=10^4$ and the flow past a wind turbine at $Re_c=103600$ were reduced by $25\%$ and $29 \%$, respectively. \\
Additionally, we have proposed to merge the viscous/turbulent-inviscid methodology with the $P$-adaptation strategy introduced in \cite{adaptation_paper}.
By combining the viscous/turbulent-inviscid coupling with $P$-adaptation, we achieve a significant reduction in computational costs without compromising the results accuracy. The computational times for the NACA0012 airfoil at $Re= 10^4$ were remarkably reduced by $41\%$. Similarly, for the flow past a wind turbine at $Re_c=103600$, we have achieved a significant reduction in computational times, with a decrease of $45\%$.

\section*{Acknowledgments}
 Kheir-Eddine Otmani aknowledges the Grant 072 Bis/PG/Espagne/2020-2021 of Ministère de l'Enseignement Supérieur et de la Recherche Scientifique, République Algérienne Démocratique et Populaires.
Gonzalo Rubio and Esteban Ferrer acknowledge the funding received by the Grant DeepCFD (Project No. PID2022-137899OB-I00) funded by MICIU/AEI/10.13039/501100011033 and by ERDF, EU.  
Esteban Ferrer would like to thank the support of
Agencia Estatal de Investigación (for the grant "Europa Excelencia 2022" Proyecto EUR2022-134041/AEI/10.13039/501100011033) y del Mecanismo de Recuperación y Resiliencia de la Unión Europea, and the Comunidad de Madrid and Universidad Politécnica de Madrid for the Young Investigators award: APOYO-JOVENES-21-53NYUB-19-RRX1A0.  
Finally, all authors gratefully acknowledge the Universidad Politécnica de Madrid (www.upm.es) for providing computing resources on Magerit Supercomputer.

\appendix

\section{Compressible Navier-Stokes}
\label{sec:cNS}
 In this work we solve the 3D Navier-Stokes equations for laminar cases and we supplement the equations with the Vreman LES model for turbulent flows. The 3D Navier-Stokes equations when including the  Vreman model can be compactly written as:
%
\begin{equation}
\boldsymbol{u}_t+ \nabla \cdot {{F}}_e = \nabla\cdot{F}_{v,turb},
\label{eq:compressibleNScompact}
\end{equation}
where $\boldsymbol{u}$ is the state vector of large scale resolved conservative variables $\boldsymbol{u} = [ \rho , \rho v_1 , \rho v_2 , \rho v_3 , \rho e]^T$, ${F}_e$ are the inviscid, or Euler fluxes,
\begin{equation}
{F}_e = \left[\begin{array}{ccc} \rho v_1 & \rho v_2 & \rho u_3 \\
                                                                                \rho v_1^2 + p & \rho v_1v_2 & \rho v_1v_3 \\
                                                                                	\rho v_1v_2 & \rho v_2^2 + p & \rho v_2v_3 \\
                                                                                	\rho v_1v_3 & \rho v_2v_3 & \rho v_3^2 + p \\
                                                                                	\rho v_1 H & \rho v_2 H & \rho v_3 H
\end{array}\right],
\end{equation}
where $\rho$, $e$, $H=E+p/\rho$, and $p$ are the large scale density, total energy, total enthalpy and pressure, respectively, and $\vec{v}=[v_1,v_2,v_3]^T$ is the large scale resolved velocity components. Additionally, ${F}_{v,turb}$ defines the viscous and turbulent fluxes,
\begin{equation}
{F}_{v,turb}= \left[\begin{array}{ccc}0 & 0 & 0\\
 \tau_{xx} & \tau_{xy} & \tau_{xz} \\
 \tau_{yx} & \tau_{yy} & \tau_{yz} \\
 \tau_{zx} & \tau_{zy} & \tau_{zz} \\
 \sum_{j=1}^3 v_j\tau_{1j} + \kappa T_x& \sum_{j=1}^3 v_j\tau_{2j} + \kappa T_y& \sum_{j=1}^3 v_j\tau_{3j} + \kappa T_z
\end{array}\right],
\label{eq:viscousfluxes}
\end{equation}
where $\kappa$ is the thermal conductivity, $T_x, T_y$ and $T_z$ denote the temperature gradients and the stress tensor $\boldsymbol{\tau}$ is defined as $\boldsymbol{\tau} = (\mu+\mu_t)(\nabla \vec{v} + (\nabla \vec{v})^T) - 2/3(\mu+\mu_t) \boldsymbol{I}\nabla\cdot\vec{v}$, with $\mu$ the dynamic viscosity, $\mu_t$ the turbulent viscosity (in this work defined through the Vreman 
model) and $\boldsymbol{I}$ the three-dimensional identity matrix. Note that when solving laminar flows, it suffices to set $\mu_t=0$ and re-interpret the large scale resolved components as the only components (there is no under-resolved components).
The dynamic turbulent viscosity using the Vreman \cite{Vreman_2004} model is given by: 
\begin{equation}
\begin{split}
    &\mu_t = C_v \rho\sqrt{\frac{B_\beta}{\alpha_{ij}\alpha_{ij}}},\\
    &\alpha_{ij} = \frac{\partial u_j}{\partial x_i},\\
    &\beta_{ij} = \Delta^2\alpha_{mi}\alpha_{mj},\\
    &B_\beta = \beta_{11}\beta_{22} -\beta_{12}^2 +\beta_{11}\beta_{33} -\beta_{13}^2 +\beta_{22}\beta_{33} -\beta_{23}^2,
\end{split}
\label{eq-iLES:LES_vreman}
\end{equation}

\noindent where $C_v=0.07$ is the constant of the model. 
The Vreman LES model adjusts the model parameters based on the local flow characteristics and automatically reduces the turbulent viscosity in laminar, transitional, and near-wall regions allowing to capture the correct physics.

\section{Spatial discretisation: discontinuous Galerkin}
\label{sec:dg}

HORSES3D discretises the Navier-Stokes equations using the Discontinuous Galerkin Spectral Element Method (DGSEM), which is a particularly efficient nodal version of DG schemes \cite{2009:Kopriva}. For simplicity, here we only introduce the fundamental concepts of DG discretisations. More details can be found in \cite{MANZANERO2020109241,FerrerJCP}.

%
The physical domain is tessellated with non-overlapping curvilinear hexahedral elements, $e$, which are geometrically transformed to a reference element, $el$. This transformation is performed using a polynomial transfinite mapping that relates the physical coordinates $\vec{x}$ and the local reference coordinates $\vec{\xi}$. The transformation is applied to \eqref{eq:compressibleNScompact} resulting in the following:

\begin{equation}
J \boldsymbol{u}_t  + \nabla_\xi\cdot{F}_e = \nabla_\xi\cdot{F}_{v,turb},
\label{eq:compressibleNScompact_transformed}
\end{equation}
where $J$ is the Jacobian of the transfinite mapping, $\nabla_\xi$ is the differential operator in the reference space and ${F}$ are the contravariant fluxes \cite{2009:Kopriva}. 

To derive DG schemes, we multiply \eqref{eq:compressibleNScompact_transformed} by a locally smooth test function $\phi_j$, for $0\leq j\leq P$, where $P$ is the polynomial degree, and integrate over an element $el$ to obtain the weak form
\begin{equation}\label{eq::NS2}
\int_{el}J \boldsymbol{u}_t\phi_j+\int_{el} \nabla_\xi\cdot{F}_e\phi_j  =\int_{el} \nabla_\xi\cdot{F}_{v,turb}\phi_j.
\end{equation}
We can now integrate by parts the term with the inviscid fluxes, ${F}_e$, to obtain a local weak form of the equations (one per mesh element) with the boundary fluxes separated from the interior
\begin{equation}\label{eq::NS3}
\int_{el}J \boldsymbol{u}_t\phi_j +  \int_{\partial el} {F}_e\cdot\hat{\mathbf{n}}\phi_j-\int_{el} {F}_e\cdot\nabla_\xi\phi_j
=\int_{el} \nabla_\xi\cdot{F}_{v,turb}\phi_j,
\end{equation}
where $\hat{n}$ is the unit outward vector of each face of the reference element ${\partial el}$. 
We replace discontinuous fluxes at inter--element faces by a numerical inviscid flux, ${F}_{e}^{\star}$, to couple the elements, 
\begin{equation}\label{eq::NS4}
\int_{el}J \boldsymbol{u}_t\cdot\phi_j + \int_{\partial el} {{F}_{e}^{\star}}\cdot\hat{\mathbf{n}}\phi_j-\int_{el} {F}_e\cdot\nabla_\xi\phi_j
=\int_{el} \nabla_\xi\cdot{F}_{v,turb}\phi_j.
\end{equation}
This set of equations for each element is coupled through the Riemann fluxes ${F}_{e}^{\star}$, which governs the numerical characteristics, see for example the classic book by Toro \cite{toro2009riemann}. Note that one can proceed similarly and integrate the viscous terms by parts (see, for example, \cite{Unified,ferrer2016a,FERRER2011224,FERRER20127037}). The viscous terms require further manipulations to obtain usable discretisations (Bassi Rebay 1 and 2 or Interior Penalty). Viscous and turbulent terms are discretised following the same spatial discretisation and in this work we retain the Bassi Rebay 1 scheme. For simplicity, here we retain the volume form:
\begin{equation}\label{eq::NS5}
\int_{el}J \boldsymbol{u}_t\cdot\phi_j + \int_{\partial el} \underbrace{{{F}_{e}^{\star}}\cdot\hat{\mathbf{n}}}_\text{Convective fluxes}\phi_j-\int_{el} {F}_e\cdot\nabla_\xi\phi_j
=\int_{el} (\underbrace{\nabla_\xi\cdot{F}_{v,turb}}_\text{Viscous and Turbulent fluxes})\cdot\phi_j.
\end{equation}


The final step, to obtain a usable numerical scheme, is to approximate the numerical solution and fluxes by polynomials (of order $P$) and to use Gaussian quadrature rules to numerically approximate volume and surface integrals. In HORSES3D we allow for Gauss-Legendre or Gauss--Lobatto quadrature points, but we only use Gauss-Legendre in this work.  

\section{A comparison between dynamic and static clustering}  \label{staticVSdynamic}
In this section, we show the impact of employing dynamic clustering vs using static clustering. Unlike static clustering, which relies on a single flow snapshot (after the flow has been fully developed or statistically converged), dynamic clustering is performed, with predefined intervals, during the simulation. Table \ref{tab:DynamicVsStatic}, shows the ratios of dynamical and static clustering costs from the total computational time of the simulations considered in this work are presented.
\begin{table}[h!]                           
	\centering
	\caption{Comparison between dynamical and static clustering}
	\label{tab:DynamicVsStatic}
	
	\begin{tabular}{p{6cm} p{5cm} p{5cm}}
		\toprule
		Test case & Dynamical clustering time over Comp.time & Static clustering time over Comp.time \\
		\toprule
          NACA0012, $Re=10^3$, $\alpha=10^{\circ}$ & $2 \times 10^{-3}$  &  $3.7 \times 10^{-10}$  \\
          NACA0012, $Re=10^3$, $\alpha=20^{\circ}$ & $7 \times 10^{-3}$ &  $7 \times 10^{-8}$ \\
	   NACA0012, $Re=10^4$, $\alpha=10^{\circ}$ & $8 \times 10^{-3}$ & $7.4 \times 10^{-5}$  \\
	   Wind turbine, $Re_c=103600$ & $3 \times 10^{-2}$ & $1.5 \times 10^{-7}$ \\
		 
		\bottomrule
	\end{tabular}
\end{table}
These results are obtained by dividing the clustering cost for both dynamic and static cases over the total computational time of the simulations. The results reveal that dynamic clustering is costly compared to static clustering for all the considered test cases. However, let us note that performing static clustering requires the flow to be fully developed, moreover, static clustering is only valid for the flows that have time-periodic behavior whereas dynamic clustering can be applied to a more general category of fluid flows (e.g. fluids that do not have main vortex shedding frequency). Even in the cases where static clustering is applicable, we typically  collect several flow snapshots to confirm that all can be represented by a single cluster, see examples in our previous work \cite{clustering_paper}. In general, static clustering (when applicable), is more efficient while providing similar levels of accuracy, see also \cite{adaptation_paper}.


\end{document}